\documentclass{svproc}
\usepackage{url}

\usepackage[dvipdfmx]{graphicx} 
\usepackage{amsmath,amssymb,bm} 

\begin{document}
\mainmatter              
\title{Wavepacket Approach for Spin Transport in Zigzag Spin Chain}
\titlerunning{Wavepacket Approach for Spin Transport in Zigzag Spin Chain}  
%
\author{Hiroaki Onishi}
\authorrunning{H. Onishi} 
%
%
\institute{Advanced Science Research Center, Japan Atomic Energy Agency,\\ Tokai, Ibaraki 319-1195, Japan\\
\email{onishi.hiroaki@jaea.go.jp}
}

\maketitle              

\begin{abstract}
We study the spin transport property of a spin nematic liquid in a frustrated zigzag spin chain in a magnetic field
from the perspective of the time evolution of wavepackets
by a time-dependent density-matrix renormalization group method.
We use the periodic boundary condition
to avoid an edge-induced magnetization structure in the open boundary condition.
We find that at the saturation,
a magnon-pair wavepacket of momentum $k=\pi$ stays localized,
since the gapless dispersion due to the antiferro-quadrupole quasi-long-range order
has a rather flat structure,
indicating zero propagation velocity.
\keywords{spin nematic liquid, spin transport, wavepacket dynamics, time-dependent density-matrix renormalization group}
\end{abstract}

\section{Introduction}

Frustrated quantum magnets have provided
a fertile playground realizing novel spin states that have no classical counterpart.
In addition to aspects of quantum magnetism,
spin transport phenomena mediated by peculiar quantum magnetic excitations have recently attracted growing interest
in spintronics
\cite{Hirobe2017,Bertini2021,Onishi2019,Hirobe2019,Onishi2022,Koga2020,Minakawa2020,Chen2021},
from a fundamental viewpoint and potential use for application.

As a marked example,
we have focused on a spin nematic liquid realized in a zigzag spin chain in a magnetic field,
which is characterized by the formation of a two-magnon bound state
\cite{Kecke2007,Hikihara2008,Sudan2009}.
We have studied magnetic and transport properties,
such as excitation spectra and spin current correlations,
by using density-matrix renormalization group (DMRG) and numerical diagonalization methods
\cite{Onishi2015a,Onishi2015b,Onishi2018,Onishi2019,Onishi2022}.
We have argued that the low-energy excitation is governed by bound magnon pairs,
so that magnon pairs would carry spin current.

In this paper,
we study the spin transport property of the spin nematic liquid in the zigzag spin chain
from the perspective of the wavepacket dynamics
by a time-dependent DMRG method.
We use the periodic boundary condition rather than the open boundary condition usually used in DMRG calculations
to avoid an edge-induced magnetization structure
that disturbs the wavepacket propagation.
We create a magnon-pair wavepacket in a spin nematic regime,
and examine how the wavepacket propagates with keeping its coherence as the time evolves.

\section{Model and Numerical Method}

We consider a spin-1/2 $J_{1}$-$J_{2}$ Heisenberg model on a one-dimensional chain of $N$ sites,
described by
\begin{equation}
  H =
  J_{1} \sum_{i} \bm{S}_{i} \cdot \bm{S}_{i+1}
  + J_{2} \sum_{i} \bm{S}_{i} \cdot \bm{S}_{i+2}
  - h \sum_{i} S_{i}^{z},
\label{eq: H}
\end{equation}
where $\bm{S}_{i}$ are spin-1/2 operators at site $i$,
$J_{1}$ ($<$0) is the ferromagnetic interaction between nearest neighbors,
$J_{2}$ ($>$0) is the antiferromagnetic interaction between next-nearest neighbors,
and $h$ is the magnetic field.
In this paper, we fix $J_{1}=-1$, $J_{2}=1$, and $N=64$,
while $h$ is set to give a total magnetization $M=\sum_{i}S_{i}^{z}$.
Note that $M$ is conserved
and used to block-diagonalize the Hamiltonian.
We use the unit such that $\hbar=1$,
and the time is measured in units of $\hbar/J_{2}$.

\begin{figure}[t]
\centering
\includegraphics[scale=0.5]{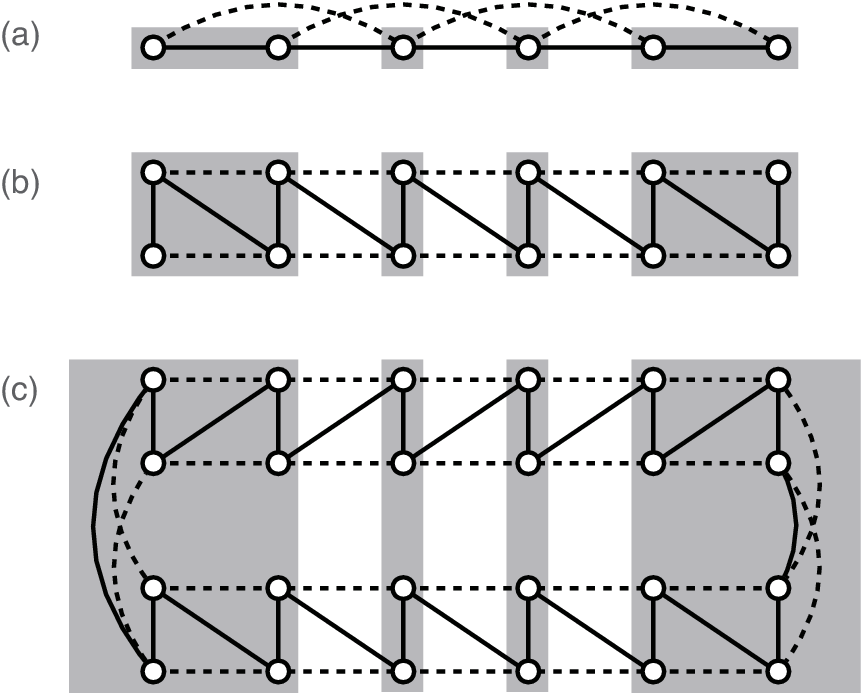}
\caption{
Superblock configuration of zigzag chain in different representations.
Solid lines denote nearest-neighbor bonds,
dashed lines denote next-nearest-neighbor bonds,
and shaded rectangles represent the superblock
consisting of left and right blocks and two (super)sites.
(a) Chain representation.
(b) Two-leg ladder representation,
where odd and even sites are paired to form a supersite.
(c) Four-leg ladder representation,
where two two-leg ladders are arranged and the periodic boundary condition is adopted.
}
\label{Fig_superblock}
\end{figure}

We first obtain the ground state
$\vert \psi_{\mathrm{G}} \rangle$
by a static DMRG method
with the use of the finite-system algorithm
\cite{White1992}.
Note that we adopt a rod-shaped superblock under the open boundary condition
to treat both open and periodic chains,
as shown in Fig.~\ref{Fig_superblock}.
Then, we prepare an initial state at time $t=0$
by creating a magnon-pair wavepacket
centered at position $j_{0}$ with mean momentum $k_{0}$,
\begin{equation}
  \vert \psi(0) \rangle =
  A \sum_{j} \mathrm{e}^{-(j-j_{0})^2/2\sigma^{2}}
  \,
  \mathrm{e}^{-\mathrm{i}k_{0}(j-j_{0})} S_{j}^{-}S_{j+1}^{-}
  \vert \psi_{\mathrm{G}} \rangle,
\end{equation}
where $\vert \psi(t) \rangle$ is the wavefunction at time $t$,
$\sigma$ is the width of the wavepacket in the real space at initial time,
while $1/\sigma$ is the width in the momentum space,
and $A$ is a normalization factor.
When $\sigma=0$, we operate $S_{j}^{-}S_{j+1}^{-}$ just at $j_{0}$.
After that, the time evolution of the wavefunction is computed
by an adaptive time-dependent DMRG method
\cite{Daley2004,White2004}.

Let us explain time-dependent DMRG techniques
used to analyze the zigzag spin chain.
In general, the time evolution of the wavefunction is described by
the formal solution of the time-dependent Schr\"{o}dinger equation as
\begin{equation}
  \vert \psi(t) \rangle =
  \exp(-\mathrm{i}Ht)
  \vert \psi(0) \rangle.
\end{equation}
By the second-order Suzuki-Trotter decomposition,
the time-evolution operator is expressed by
the product of local time-evolution operators with a small time step $t/2n$,
set to be $0.02$ in the present calculations,
as
\begin{align}
  \exp(-\mathrm{i}Ht)
  =&
  \left[
  \exp\left(-\mathrm{i}H_{1}t/2n\right)
  \exp\left(-\mathrm{i}H_{2}t/2n\right)
  \cdots
  \exp\left(-\mathrm{i}H_{\tilde{N}}t/2n\right)
  \right.
\nonumber \\
  &
  \left.
  \times
  \exp\left(-\mathrm{i}H_{\tilde{N}}t/2n\right)
  \cdots
  \exp\left(-\mathrm{i}H_{2}t/2n\right)
  \exp\left(-\mathrm{i}H_{1}t/2n\right)
  \right]^{n},
\end{align}
where the total Hamiltonian is decomposed into
$\tilde{N}$ terms of local Hamiltonians as $H=\sum_{i}H_{i}$.
The local time-evolution operator can be sequentially multiplied to the wavefunction
without the truncation error through the left-to-right and right-to-left sweep procedure
if there are no long-range terms beyond adjacent blocks,
since we can use untruncated bases of the single site for the calculation of the matrix multiplication.
Note that the value of $\tilde{N}$ depends on how we setup the superblock configuration.
For a chain representation of the zigzag chain
in Fig.~\ref{Fig_superblock}(a),
the local Hamiltonian is given by
\begin{equation}
  H_{i} =
  J_{1} \bm{S}_{i} \cdot \bm{S}_{i+1}
  + J_{2} \bm{S}_{i} \cdot \bm{S}_{i+2}
  - h (S_{i}^{z}+S_{i+1}^{z})/2,
\end{equation}
and $\tilde{N}=N-1$.
Hence, the second term acts beyond adjacent blocks,
and the multiplication of the local time-evolution operator causes the truncation error.
A practical way to avoid the truncation error is
to consider a two-leg ladder representation
in Fig.~\ref{Fig_superblock}(b).
The local Hamiltonian is explicitly given by
\begin{align}
  H_{i}
  =&
  J_{1} \bm{S}_{2i+1} \cdot \bm{S}_{2i+2}
  + J_{1} \bm{S}_{2i+2} \cdot \bm{S}_{2i+3}
  + J_{1} \bm{S}_{2i+3} \cdot \bm{S}_{2i+4}
\nonumber \\
  &
  + J_{2} \bm{S}_{2i+1} \cdot \bm{S}_{2i+3}
  + J_{2} \bm{S}_{2i+2} \cdot \bm{S}_{2i+4}
\nonumber \\
  &
  - h (S_{2i+1}^{z}+S_{2i+2}^{z}+S_{2i+3}^{z}+S_{2i+4}^{z})/2,
\end{align}
and $\tilde{N}=N/2-1$.
It includes only terms within adjacent blocks,
and thus the local time-evolution operator can be multiplied properly without the truncation error.
We note that a supersite consists of two sites,
so that the number of states per supersite becomes large as $2^2=4$,
indicating the increase of computational cost for each matrix operation due to the large matrix dimension.
Moreover,
as shown in Fig.~\ref{Fig_superblock}(c),
it is useful to consider a four-leg ladder representation
to adopt the periodic boundary condition
\cite{Tzeng2012}.
Here, a supersite has four sites,
and the number of states per supersite increases to $2^4=16$.
We mention that from the viewpoint of the matrix dimension,
this four-leg ladder system corresponds to a one-dimensional two-orbital Hubbard model,
for which we have studied the real-time dynamics
by successfully applying the adaptive time-dependent DMRG method
\cite{Onishi2010,Onishi2013}.

\begin{figure}[t]
\centering
\includegraphics[scale=0.55]{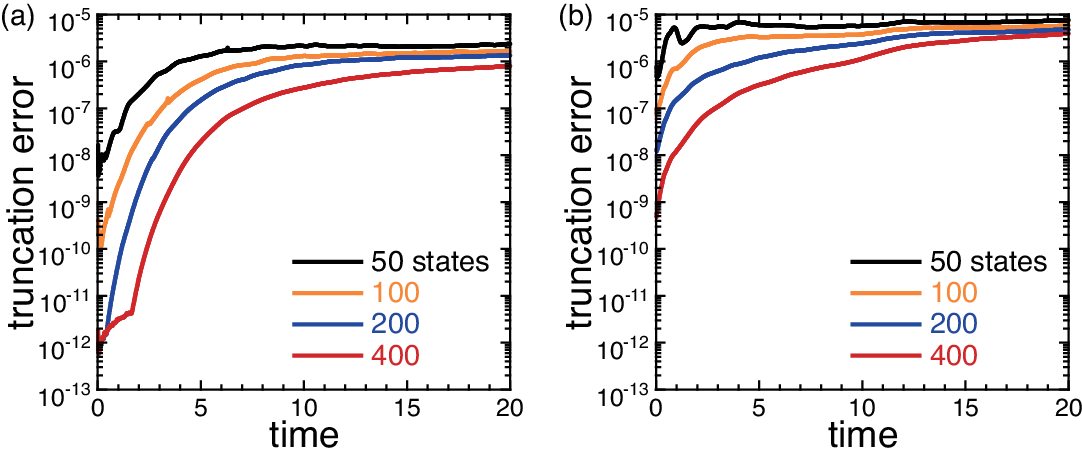}
\caption{
Truncation error as a function of time in
(a) open and
(b) periodic boundary conditions
for several values of the number of states kept.
$J_{1}=-1$, $J_{2}=1$, $N=64$, $M=24$, and $\sigma=0$.
}
\label{Fig_truncation}
\end{figure}

Note that the superblock configuration of the four-leg ladder representation resembles
that of the two-leg ladder representation
except for the number of legs and the site connection just at edges.
However,
we should notice that two cross sections of the two-leg ladder appear when we cut the system
in the four-leg ladder representation.
The states at these cross sections are nearly independent when the system is large.
Thus, the four-leg ladder representation for the periodic system is computationally expensive
due to the increase of the number of states required to keep the truncation error small.
In Fig.~\ref{Fig_truncation},
we show the truncation error for $N=64$ and $M=24$.
Keeping up to $400$ states,
the truncation error in the open boundary condition is $2 \times 10^{-12}$ in the ground state
and kept below $10^{-6}$ during the time evolution,
while that in the periodic boundary condition
is $10^{-8}$ in the ground state and kept below $4\times 10^{-6}$ during the time evolution.
For the saturation $M=32$,
the truncation error is below $10^{-10}$ even with $100$ states
in the both boundary conditions.

\section{Numerical Result}

\begin{figure}[t]
\centering
\includegraphics[scale=0.55]{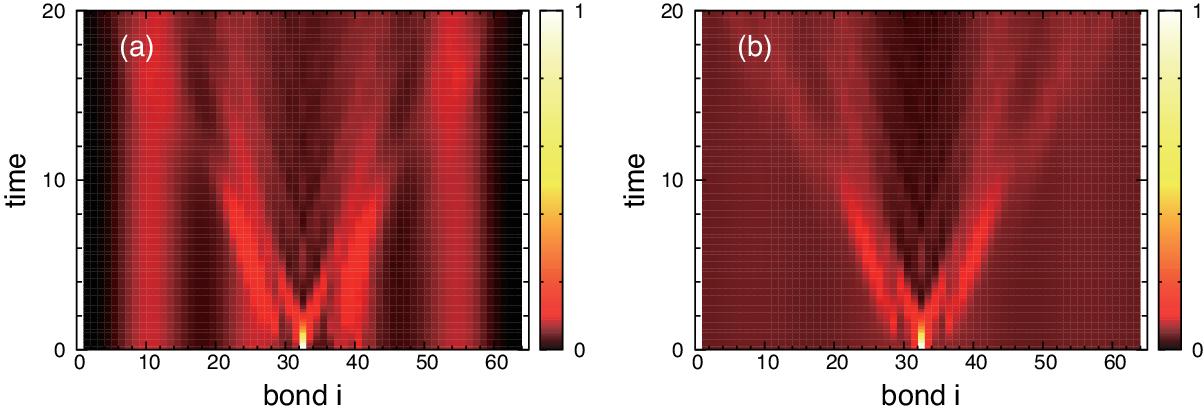}
\caption{
Time evolution of magnon-pair density $N^{--}(i,t)$ in
(a) open and
(b) periodic boundary conditions.
$J_{1}=-1$, $J_{2}=1$, $N=64$, $M=24$, and $\sigma=0$.
Here we keep $400$ states.
}
\label{Fig_ndd_obc-pbc}
\end{figure}

To clarify how the magnon-pair wavepacket propagates,
we investigate the time evolution of the magnon-pair density, defined by
\begin{equation}
  N^{--}(i,t)=\langle S_{i}^{-}S_{i+1}^{-} S_{i}^{+}S_{i+1}^{+} \rangle_{t},
\end{equation}
where $\langle\cdots\rangle_{t}$ is the expectation value using the wavefunction at time $t$.
Note that $S_{i}^{-}S_{i+1}^{-}$ and $S_{i}^{+}S_{i+1}^{+}$ are
creation and annihilation operators of a magnon pair, respectively.
Figure~\ref{Fig_ndd_obc-pbc}(a) represents $N^{--}(i,t)$
in the open boundary condition.
Here, $N=64$ and $M=24$, i.e.,
eight spins are flipped down from the fully polarized spin-up state.
We find that
eight down spins are paird to form four magnon pairs in the ground state,
while the four magnon pairs sit separately in the open chain.
Thus, we have an edge-induced magnetization structure.
After creating a magnon-pair wavepacket at the chain center,
it expands left and right in the chain as the time evolves,
whereas the originally existing four magnon pairs stay localized.
We see that left and right wavefronts move at some velocity,
and eventually they each hit an originally existing magnon pair.
There, they are partly transmitted and partly reflected due to a barrier.
Such an open-boundary effect is removed in the periodic boundary condition,
as shown in Fig.~\ref{Fig_ndd_obc-pbc}(b).
The ground state is uniform, i.e.,
magnon pairs distribute uniformly in the periodic chain,
so that the magnon-pair wavepacket propagates without the disturbance of
the edge-induced magnetization structure.
Therefore, hereafter, we adopt the periodic boundary condition.

\begin{figure}[t]
\centering
\includegraphics[scale=0.55]{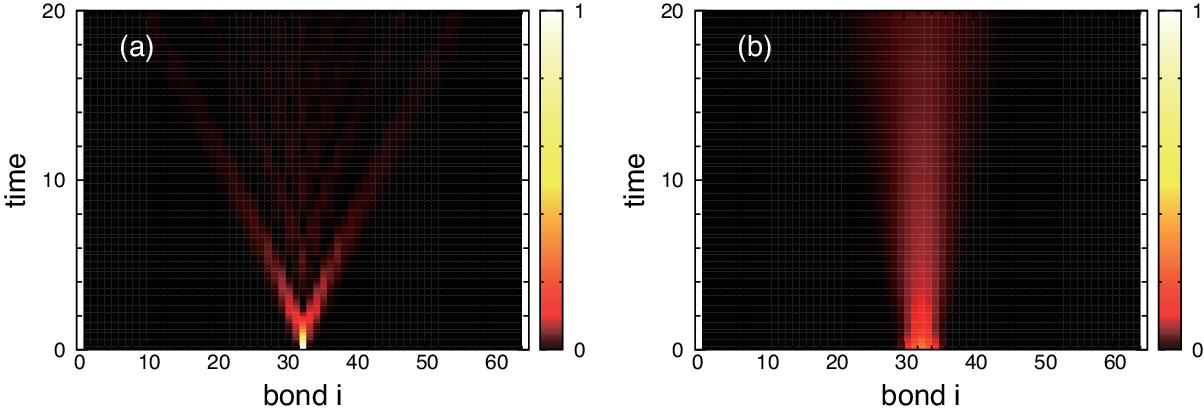}
\caption{
Time evolution of magnon-pair density $N^{--}(i,t)$ with
(a) $\sigma=0$ and
(b) $k_{0}=\pi$ and $\sigma=2$.
$J_{1}=-1$, $J_{2}=1$, $N=64$, and $M=32$.
Here we adopt the periodic boundary condition
and keep $100$ states.
}
\label{Fig_ndd_s0-k1s2}
\end{figure}

In Fig.~\ref{Fig_ndd_s0-k1s2}(a),
we show $N^{--}(i,t)$ with $\sigma=0$ at the saturation.
We observe that wavefronts move left and right at a constant velocity,
implying that magnon pairs propagate with keeping its coherence.
Note here that for $\sigma=0$,
all of momenta are equally involved in the wavepacket.
Each component propagates with its velocity,
determined by the slope of the dispersion of the quadrupole excitation spectrum.
The velocity of the wavefront in Fig.~\ref{Fig_ndd_s0-k1s2}(a)
corresponds to the maximum among all components.
However, for finite $\sigma$,
the range of relevant momenta is limited around the mean momentum $k_{0}$
within a width $1/\sigma$.
In Fig.~\ref{Fig_ndd_s0-k1s2}(b),
we show $N^{--}(i,t)$ with $k_{0}=\pi$ and $\sigma=2$ at the saturation.
We clearly see that the wavepacket stays localized.
This is because the quadrupole excitation spectrum has a gapless mode
due to the antiferro-quadrupole quasi-long-range order,
and the gapless dispersion has a rather flat structure
\cite{Kecke2007,Onishi2018},
indicating zero propagation velocity.

\section{Summary}

We have studied the time evolution of the magnon-pair wavepacket
to gain an insight into the spin transport property of the spin nematic liquid in the zigzag spin chain.
We have observed that magnon pairs propagate with keeping its coherence.
The magnon-pair wavepacket of $k_{0}=\pi$ is found to stay localized
due to the flat structure of the gapless dispersion of the quadrupole excitation spectrum.

\section*{Acknowledgement}

Computations were performed on supercomputers
at the Japan Atomic Energy Agency
and at the Institute for Solid State Physics, the University of Tokyo.
This work was in part supported by JSPS KAKENHI Grant Nos.~19K03678 and 23K03331.


\end{document}